# Co-Designing a Medication Notification Application with Multi-Channel Reminders

## Completed research paper

### Nawal Chanane
School of Engineering, Computer and Mathematical Sciences
Auckland University of Technology
Auckland, New Zealand
Email: nawal.chanane@aut.ac.nz

### Farhaan Mirza
School of Engineering, Computer and Mathematical Sciences
Auckland University of Technology
Auckland, New Zealand
Email: farhaan.mirza@aut.ac.nz

### M. Asif Naeem
School of Engineering, Computer and Mathematical Sciences
Auckland University of Technology
Auckland, New Zealand
Email: muhammad.asif.maeen@aut.ac.nz

## Abstract

Evidence suggests that medication adherence applications (apps) are one of the most effective methods to remind patients to take medication on time. Reminders via apps are overwhelming today, consumers discard using them after a brief period of initial usage, eventually becoming unfavourable towards them and not using them at all. This study aims to qualitatively determine the key features and design of medication reminder apps that facilitate or disrupt usage from the users' perceptive. Three focus groups were conducted with participants aged between 15 and 65+ (N= 12). The participants evaluated a smart medication reminder prototype, then sketched and discussed their thoughts and perceptions within the group. Participants identified, 1) Multi-channel reminders, 2) Medication intake acknowledgement for reporting and 3) Seamless addition of medications and associated reminders as important elements. Understanding consumers' needs and concerns will inform the future development of medication reminder apps that are acceptable and valuable to consumers.

# 1 Introduction

Patient's adherence to taking medications accurately and remembering to take their prescribed medications has become an ever-growing issue between doctors and patients (Brown and Bussell 2011). Medication app reminders help users to remember to take their medication on time. However, in spite of the growing adoption of the mobile health (mHealth) apps, their sustainability and user's low retention rate concerns the researchers and few studies addressed this matter (Tison et al. 2018). The low retention rate is caused by the incorrect use of the applications, thus to understand benefits from the technology adoptions is required (Gücin and Berk 2015). The literature suggested that, while initial adoption of technology is related to the cost of exploring an innovation such as perceptions of compatibility or visibility, long-term engagement is primarily driven by rational consideration (Agarwal and Prasad 1997). Therefore, users may become attracted to the availability of new technology. However, this attractiveness may be temporary unless they develop forms of using this technology to gain its related benefits over past practices.

This paper focuses on the design of a Minimum Viable Product (MVP) medication reminder app, following a qualitative approach with users as the first beneficiaries of the app. This research is foreseen to inform future mHealth attempts and assist providers to better guide app designers and developers on user-centred design that serves the users' needs. The contributions of this study are as follows:

- Collects end-user's feedback and identify what is considered main elements in a medication reminder app from the user's perception as the first consumers.
- Captures the key elements to overcome the challenge of med reminder apps usage continuity.
- Build an MVP through an iterative design process.

The remainder of this paper is organized as follows: Section 2, presents related work. Then, Section 3 describes the methodology, including sample and participants selection, procedure, initial prototype and data collection. Section 4, introduces results, discusses the findings and presents the output of the analysis in the form of an MVP. Finally, Section 5, drawn the conclusion and proposed future work.

# 2 Related Work

## 2.1 mHealth Apps as Medication Reminders

mHealth can be used as a powerful health-behavior change tool for health prevention and ubiquitous medication self-management that can be carried by a person and has advanced computational capacity (Schnall et al. 2017). In particular, medication adherence (MA) apps are useful tools for helping patients take their medications as prescribed (Park et al. 2019). Moreover, they are being increasingly used as part of mHealth approaches to provide low-cost and sustainable healthcare to patients with chronic illnesses. However, mHealth apps are continuing to multiply with little evidence for their effectiveness and little support for understanding how to best design them (Mccurdie et al. 2012). Many current mHealth interventions are designed on the basis of existing healthcare-system constructs and may not be as effective as those that involve end-users in the design process (Schnall et al. 2017). Apps need to be created with adequate consideration of the needs of their intended users, so that they are both easy to use and perceived as useful (Joyce 2019).

The prior studies have incorporated different evaluation methods to assess app features, including using proprietary rating systems (Haase et al. 2017) and user testing (Schnall et al. 2017). Furthermore, evaluations of app reviews have been limited, although app reviews posted by targeted users are publicly accessible and add to a valuable generated pool of data that has not been fully utilized (Zusman 2018). Together, the continuously growing number of mobile phone users (Statista 2018) and the greater recognition of the problem of MA in recent years (Zusman 2018) requires an update of these studies' approaches.

## 2.2 Co-Design via Focus Groups

As mHealth technologies expand the design of effective tools is becoming increasingly important (Fiordelli et al. 2013). In particular, focus groups are useful for identifying the requirements of targeted consumers, and a meaningful discussion of basic ideas, concepts and designs is possible within such groups. Discussions can reveal conflicts and contradictions among participants and produce an overview of different opinions (Kitzinger 1994).Focus group discussions are frequently used as a qualitative approach to gain an in-depth understanding of actual usage.

This method aims to obtain data from a purposely selected group of individuals rather than from a statistically representative sample of a broader population. Focus group discussions are appropriate for the generation of new ideas formed within a social context than one-to-one interviews (O.Nyumba et al. 2018). Moreover, their results can be extremely valuable, and the insights obtained can lead to an improved product or service that consumers really want and will purchase.

## 3 Methodology

For an effective user-experience the key considerations are design, development iterations, and evaluation (Traynor et al. 2017). If a study identifies several issues with a prototype, it can then be fixed by a new design. Based on the MA management requirements elicited in the previous research (Chanane et al. 2019), a mobile app prototype is developed, that is simple to use and serves patients and clinicians by keeping a record of medication status and the progress of medication intake. In Gracey et al. (2018) research, they discussed how likely an app is to be successful by proposing ways to foster creativity and to further develop ideas for its design and features. Moreover, they suggested that developers and users of the app are to be consulted, and that their feedback should be incorporated into the design.

The aim of co-design is to co-create solutions to problems using the knowledge and expertise of those who have authentic experience of the issue or need being investigated (Labattaglia 2019). Co-design places the 'user' as 'an expert' of their experiences (Burkett 2016). It involves designing an MVP through a co-design focus group session that can sustain beyond the scientific evaluation. To increase the impact of technology and avoid the ad-hoc style developments that neglect the needs of the target users, this research adopted the CeHRes framework for eHealth design and development (van Gemert-Pijnen et al. 2011). This framework requires continuous iteration and evaluation with user that spans the entire development. Following a formative evaluation, in this research, medication mHealth users were consulted through three co-design focus group sessions. Each session was considered an iteration for feedback and evaluation. The outcomes of each iteration were added to the design to obtain feedback from the next iteration, see Figure 1. The process included users who brought a deep understanding of their contexts and needs, and the opportunities arising from these were explored. Their perceptions helped the designer identify what needs to be changed to improve the interface. It is anticipated that findings from this research will produce a user-centred medication smart-reminder MVP.

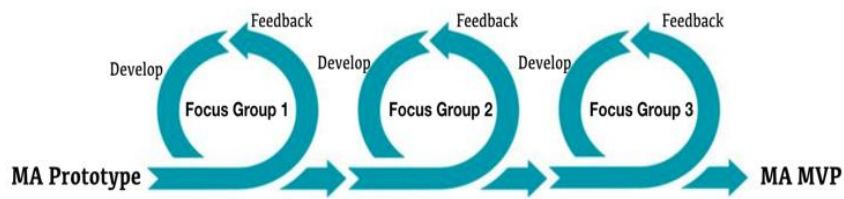

*Figure 1: Focus groups co-design iterations for MAMA app*

### 3.1 Sample Selection and Participants

Purposive sampling was used in this study (Martin 1996). Participants for the focus groups were either previous participants from the requirements elicitation phase (Chanane et al. 2019), who agreed to be contacted to participate. Or from personal contacts who showed interest on participating as users of medication reminder apps and/or consumers of medication. To get a well-shaped picture about why people used or did not use medication reminder apps, both individuals with and without prior knowledge or usage of those apps were emailed an invitation.

After their response of being keen to participate, the researcher sent through the information sheet, consent form and evaluation form. They were asked to email the consent form before attending the focus group if they wish to receive the app link to download on the phone (app version for IOS users only) before they attend the session. The non-IOS users were informed to have an iPhone device provided for them during the focus group, with the app pre-installed. A total of 12 individuals, including 4 participants in each focus group keeping in mind the distribution of age range across the 4 groups, in an attempt to generate diverse feedback from each group.

### 3.2 Focus Group Sessions Procedure

After acquiring the participants consent for recording the discussion, a printed copy of the evaluation form distributed along with an iPhone for the ones who did not have an iPhone, an A3 sketch white papers, sticker notes and colored markers and candy as a pretend medication to take during the session.

The researcher was doing the role of the moderator and note taker. The researcher did not have prior knowledge about the participants' view of the medication apps nor about their health conditions. The researcher had a positive attitude towards health apps but tried to remain neutral in the conversation with the participants. Each focus group session took place in the conference room and run for 60 to 90 minutes. All participants were provided with a NZ$25 gift card and a thank you card for their time. The focus groups discussions were audio-recorded and then transcribed verbatim.

The researcher followed the session plan developed and shared with the research team prior to the focus groups. Participants were first provided with an introduction about the purpose of the study, presenting the workflow of the expected future use of the proposed app as shown in Figure 2. They were then asked about their overall understanding about health apps, their experience with medication apps, reasons of liking or disliking them. Participant freely shared their experience without prompts. Then, the researcher went through a demo of the prototype. After that, they were asked to open the app (discovery), create an account and start navigating its screens. The researcher went around the participants and answered individuals' questions. Then, the four participants were put into two groups (to allow the discussion between peers and sharing ideas) they were asked to move to the sketching part and draw/note their thought on the A3 white paper and answer the evaluation form they were given at the start of the session.

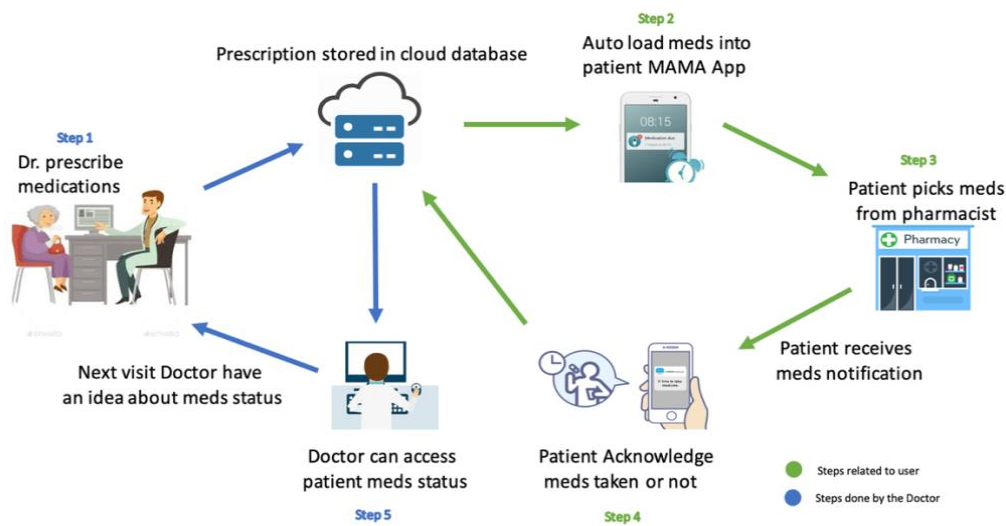

*Figure 2: Patient – Healthcare provider MAMA App usage workflow*

### 3.3    Initial Prototype

The prototype with six working screens was designed and developed in consultation with healthcare experts and health technology designers, as a result from the requirements elicitation study (Chanane et al. 2019). The prototype was given the name "MAMA" (Medication Adherence Management App). The content was designed to include the main features, that allows the users to be reminded through notifications to take their medication on a scheduled time, track stopped medication according to the medication status, share with the healthcare system. Graphics and color were included to enhance the prototype's visual appeal as shown in the medication reminders screen in Figure 3 and the list of stopped medication screen in Figure 4.

### 3.4    Data Collection

The data was collected in four ways during the focus group. The data collection began after all participants sign the consent forms and downloaded MAMA. There was an open discussion which was audio recorded when the participants started the discussion. The recording was switched on until the discussion ended. As supplementary material the sketches by participants were captured. All questions asked and answers were recorded for better understanding of their concerns and feedback. At the end of the session the data collected was: 1) evaluation form, 2) Notes of questions asked during navigation of the app, 3) sketches and 4) group discussion recording. In summary, the data gathered during the focus group session were done in four ways:

- Evaluation form: the participants were given the time to discover the prototype screens. A form was given to them to collect their feedback on the prototype sections they were also given the option of including further views and comments after each main section of the evaluation form.

- Note taking: the researcher took notes of all the questions and feedback given while participants where navigating the prototype and answering the evaluation form sections. Furthermore, the ambiguous sections were noted for detailed explanation and clarification.

- Sketching / Drawing on A3 white paper: After completing the form, the participants were asked to include all that they considered missing from the prototype to further express it as a visual representation of the prototype screens according to what they see better as users in terms of visual, content and functions.

- Group discussion audio recording: The participants drawing was discussed amang the team and audio recorded to make sure the discussion details are were not missed and to be included in the discussion section in details.

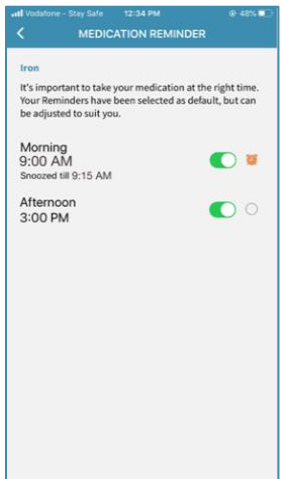 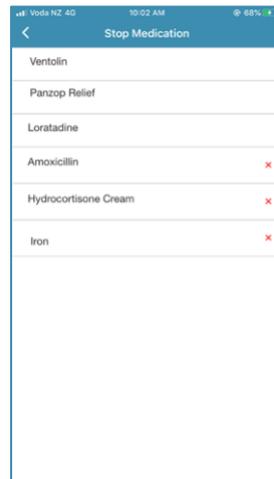 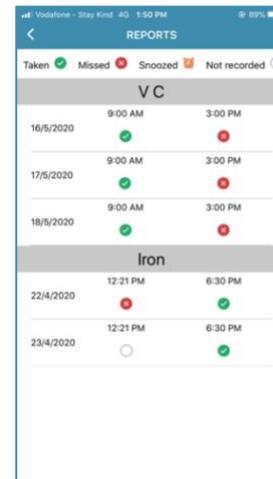

*Figure 3: Medication reminders screen.*　　*Figure 4: List of stopped medication Screen*　　*Figure 7: In app medication status report.*

## 4   Results and Discussion

Data were analysed using thematic analysis, a flexible method for qualitative analysis that allows for the identification of patterns of responding within the data (Braun and Clarke 2006). This analysis is considered suitable as it provides a detailed description of data set and generates insights into participants' perspectives on specific topic. The extracted features suggested by the users were implemented into the app prototype. The drawings on the A3 papers were compared, and evaluated according to the goal of the project, visual, navigation, purpose, content and layout.

Data were separately analysed after each focus group iteration. The output of each iteration was implemented in the prototype design to be presented in the second focus group. The intersected themes identified within the data were amended into the prototype after the third/last iteration. Themes are identified to be any subject relevant to the purpose of research that is represented by some level of patterned response within the data set (Braun and Clarke 2006). Thus, a theme was considered any subject that was discussed by participants on more than one independent occasion during the focus groups. In this paper three core themes were identified that informed our final MAMA App design that result into an MVP. The main themes were listed and discussed in detail in the coming subsections: 1) multi-channel reminders, 2) medication intake acknowledgement and reporting and 3) Connection with the health system. A visual illustration of the themes presented in Figure 5, and a sample of selected participants quotes across the three focus groups presented in Table 1.

### 4.1   Multi-channel reminders

According to Stawarz et al. (2016) reminders are important for people to remember a specific task on their busy schedule. Moreover, older medication users as they are more likely to forget when they are taking more than one medication, especially if it was taken for a long time as patient tend to forget over time. This feature has been added to accommodate a variety of users with busy schedules, who are more likely to forget their medication during the day even if they have all sorts of medication apps on their phones. The multi-channel reminders can accommodate the users who miss their dose as a result of not checking their mobiles frequently during the day but they do access their emails. And/or users who considers phone call reminder more practical than an email that can be sent to their overloaded inbox. And/or users who consider an in-app notifications one of the things they don't even look at due to the number of notifications they receive from all the apps they have on their phones. And/or users who like to be reminded multiple times to remember and take their medications. In this study the user will receive reminders from multiple channels according to user's response to the app notification. The reminders are based on a workflow in Figure 6. MAMA MVP will thus provide four channels of notifications: 1) App push notifications, 2) Email notifications, 3) SMS notifications and 4) Auto voice call reminder.

| Original Themes | Participants Quotes |
| --- | --- |
| Simplicity/ Easy to use | "It's not the sort of app that you need people to spend time using it. Reminder should be very light weight. Other apps are super crowded and annoying. Example, from a running app to do simple job to trying to sell me shoes. I don't want to play candy crash in the app!! Just keep it simple. It has a job to do." |
| | "Lots of people are not email tech, so a phone call will be more useful for them." |
| Reporting | "If I get the med from the pharmacy only, good to have it added to the app also. It doesn't have to be only when the GP prescribe it." |
| | "Good my GP will see what I am taking." |
| Integration | "The alert to GP if stops medication is important because there are medications that patient can't just stop them." |
| | "It's good to link this app to HEALTH365[1] so I can check my medication history that will make it even easier for record keeping." |
| | "It can be used in a rest home, where the nurse will receive notification." |

*Table 1. A Sample of Selected Participants Quotes Across the Three Focus Groups*

The four channels are implemented according to the notification workflow, processed based on the information provided by the user during the sign-up. By providing this, it is a step forward to design to accommodate individual needs (according to the details provided in the app) and deliver medication notifications in the way that is best and most convenient individual. However, depending on the contacts details provided by the user. To fully benefit from the email, SMS and voice call reminders the user will be required to provide the email address, caregiver contact number and phone number, otherwise the multi-channel reminders will not be fully functioning.

Health service barriers play an important role in frustrating efforts to maintain the level of adherence to a given treatment. Shubber et al. (2016) reported that forgetting medication was the most frequently cited barrier to adherence across all age groups. Challenges relating to timing of medication, including being asleep, could be overcome through text messaging, reminder devices and individual counselling that seeks to modify medication taking in a way that fits one's daily activities. Another study conducted by Shin et al. (2015) reported from interviews that participants mentioned occasions when they did not take medications on time as prescribed because they simply forgot to do so during working hours. Therefore, the proposed function, will include multiple notification channels as detailed below and presented in Figure 6:

- App push notification: this is the default and first form of reminder; the user will receive an app notification to take medication at the chosen time (the users will have the ability to choose the number of times to be reminded during the day). This is ideal way to remind as its non-intrusive and in accordance with the system wide notification preferences of the mobile phone. According to medication administration research, the users of medication have 30 minutes time up to one hour window to take their medication before it is considered late dose (Furnish et al. 2020). Therefore, the user will need to take the medication and log it in the app as taken. If the medication is not taken 30 minutes after the first dose time, the second channel (Emails) will be considered.

---

[1] HEALTH365 is an online gateway to information and services provided by the GP. And it has been developed by the same team who created the GP's information systems; ensuring services are integrated, up-to-date and secure. https://health365.co.nz/AboutUs

- Email notification: When signed up, the user will receive an email notification to the specified email address. The user will be reminded to take the medication and prompted to log it in the app. In case no acknowledgement is available for any of the scheduled medications during the day, the third channel will be applied, which is the Auto voice call.

- Auto voice call: This type of notification depends on the availability of the users' mobile number. This function will make the acknowledgement of medication intake easier especially for the ones who do not check their mobiles frequently. It is anticipated that phone calls will attract their attention, reminding them with their medication even if they did not have the chance to check the phone for any SMS or push notification. According to research concluded by Hasvold and Wootton, telephone reminders have an improved hospital attendance rate compared with SMS reminders. Therefore, this functionality will further explore and confirm if automated voice calls will impact medication intake rate in comparison with other types of notifications provided (Hasvold and Wootton 2011).

- SMS notification: In this case, the SMS notification will be directed to the caregiver number registered when signing up in the app. However, in case no caregiver mobile number is registered, a red flag will be logged in the report. A further approach will be to put a flag on the user's username in the healthcare provider's database in case multi-notification has been used more than twice during the week and still no medication has been logged.

## 4.2 Medication Intake Acknowledgment and Reporting

One of the main elements rose in the focus group sessions is the ubiquity of logging their medication intake in the app and having it handy to show their healthcare provider during their appointment. Moreover, the availability of this kind of medication reporting overtime accessible through the health system is always a bonus to their medical record and treatment progress. MAMA provides a streamlined design with a quick way of checking their medication intake status, see Figure 7. The same report can be viewed in the health system for a quick reference for the healthcare provider before meeting the patient in the clinic. In the recent study by Park et.al in (2019), they discussed a number of features available in MA apps; including reminders to take medication, reminders to refill, storing medication information and logs but none of the existing apps focused on feeding the data back to the health system.

This feature is anticipated to fill the gap of the treatment record , knowing that adhering to the medication prescribed is one of the main elements of improving health conditions (Eimear.C. et al. 2018). The following two subthemes were considered integrated to serve the same purpose of keeping track of medication:

- Medication intake report/history: This allows users to see their history of medication intake with the status of what was taken, missed or discontinued. The patient can share the report with the healthcare provider and/or pharmacist. Medication history is important for keeping an up-to-date list of all medications, including the prescribed dosage, the dosing frequency and the reason the drug was prescribed. Also, it will allow reporting the patient's dietary restrictions necessitated by a specific medication (Saljoughian 2019). Moreover, the patient's consultations with different doctors make it difficult to keep the medications list up to date because of the various health systems in primary and secondary care and in private healthcare (Kvarnstrom et al. 2018). Furthermore, being able to track treatment regimen is considered a struggle for all ages, especially with the older adults and the concern of double dosing (Schneider et al. 2019). Therefore, including this function in the app will allow the user to keep all medication history and the latest medication list at hand for his record and if needed for any emergency situation.
- Flag Stopped medication: This feature will allow users to flag discontinued medication on their app when stopped for any reason; otherwise, patients will still be taking them. Having patients changing the drug regimen or stopping any medication or replacing one medication with another, will greatly affect the treatment outcome. It will be beneficial to keep a record, so it can be shared with the healthcare provider for the possibility of reducing the frequency of administration, introducing combination medicines or even deprescribing some medication. Moreover, patients engage in their health care is a significant role for an informed and patient-centered way of treatment (Eimear.C. et al. 2018).

Therefore, patients engaging with mobile app technologies may have the potential to record all the medication intake and make the data available to GPs at the point of care. In a randomized controlled trial conducted by Vasilevskis et al. (2019), they examined the impact of medication reduction on adherence by deprescribing and it was concluded that incorporating patients' preferences into the decision-making is significant to the treatment process. Furthermore, preventing polypharmacy while keeping track of which medication is on the list to be taken and what medication has been stopped and which drug to watch out for to avoid drug interaction, which can also lead to nonadherence to the prescribed medication, are also very useful functionalities (Brinton 2018).

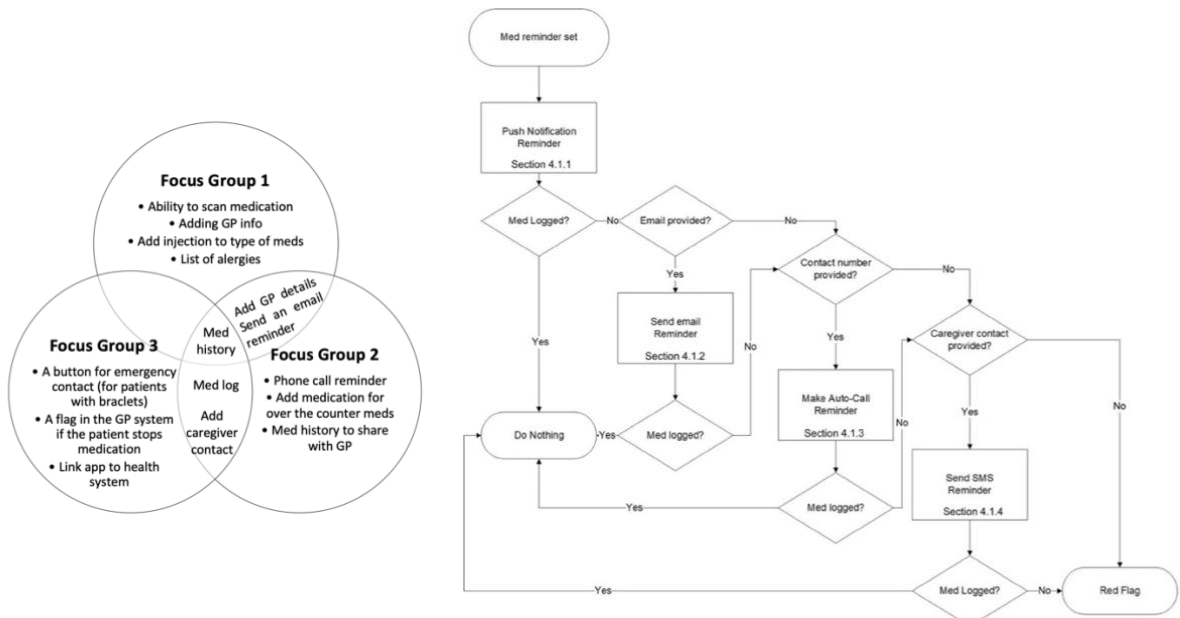

*Figure 5: High level findings from focus groups with unique and intersected themes*

*Figure 6: Multi-channel notification workflow*

## 4.3 Smart loading of Prescription to App

One of the most distinctions between the available medication reminders app and MAMA app is the auto-load of medication through a webform which could be integrated with the health system. The theme of smart loading of medication is one of the features that represents the connectivity with the health system through the app. This includes: 1) automatic loading of medication from the health system to the app users' account, 2) semi-automatic loading of medication into the app through the MAMA webform Figure 8 and 3) self-adding medication through the app using the add medication function screen shown in Figure 9. Medication adherence and polypharmacy "the use of multiple drugs or more drugs than are medically necessary" (Maher et al. 2014) are significant public-health concerns worldwide and are an important focus of integrated care. Also, older people relying upon community pharmacies frequently have difficulty managing their medications (Beuscart et al. 2019). A recently published study by Saljoughian (2019) reported that approximately 44% of men and 57% of women older than 65 years take five or more nonprescribed and/or prescribed medications per week. They concluded that technology-driven systems and electronic medical records will help prevent harmful drug effects and interactions.

Moreover, the increasing use of multiple medications has been associated with an increased risk of unfavorable drug reactions, drug to drug interactions, medication nonadherence and greater healthcare costs (Shah and Hajjar 2012). Therefore, the proposed function in MAMA is easy-to-use with providing auto-loading of medication to prevent any error on the details of Rx entry when adding medication to the app, or almost no effort from users when scanning their prescription using the feature of scan prescription shown in Figure 10, then sending it through the app to the researcher for medication loading using the webform presented in Figure 9 (the researcher is the one loading the medication for the purpose of this research only). In a real case scenario this could be done by the pharmacist when dispensing the medication to the patients or the health provider when prescribing the medications. According to research, an easy to use app with minimum purposeful functionalities, responsive and useful to the patient is more likely to be used (Kenny et al. 2014).

## 5    Conclusion

In this paper, three co-design focus group sessions were conducted with 12 end-users. The feedback gathered from each of the three iterations resulted into a medication smart reminder MVP. The participants suggested several ideas to resolve the challenge of med reminder apps usage discontinuity. Then, then focus group results were analysed to define the underlying themes. The top recommended features identified from the data collected implemented into the prototype were 1) Multi-channel reminders, 2) Medication intake acknowledgement and reporting and 3) Smart loading of prescription into the app as important elements.

The strength of this study was our inclusive sampling of the participants, who included the main users of the app, to have a complete view of the reasons for using or not using medication reminder apps. Another strength was the design of the key elements from the user's perception and building the MVP following an iterative design process among a diverse set of users. The limitation of this study is the limited number of participants which could be addressed by recruiting more users in each group. These findings showed positive perceptions of this approach and provided us with insights into possible improvements medication reminder apps. Further work is planned to trial the MAMA MVP with users as a proof of concept to evaluate its effectiveness and efficiency.

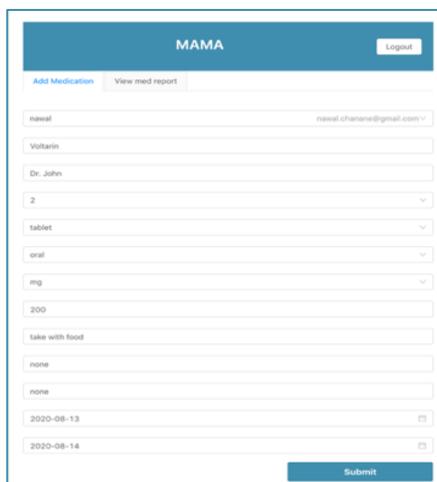
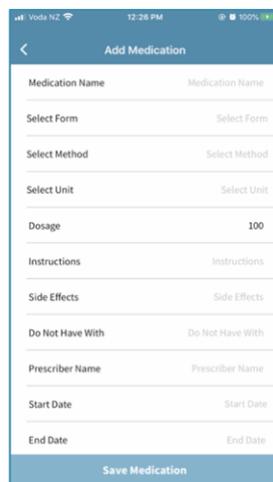
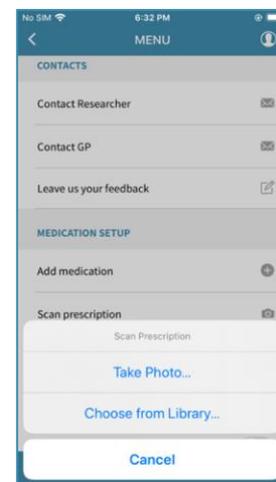

*Figure 8: Webform for auto-load of Medication*

*Figure 9: In-App add medication screen.*

*Figure 10: Scan prescription*